# Stabilization of linear carbon structures in a solid Ag nanoparticle assembly

C.S. Casari, V. Russo, A. Li Bassi, C.E. Bottani

NEMAS-Center for NanoEngineered MAterials and Surfaces and
Dipartimento di Ingegneria Nucleare,
Politecnico di Milano, Via Ponzio 34/3 I-20133 Milano, Italy

F. Cataldo

Soc. Lupi arl, Chemical Research Institute, via Casilina 1626/A, 00133 Roma, Italy

A. Lucotti, M. Tommasini, M. Del Zoppo, C. Castiglioni, G. Zerbi

NEMAS-Center for NanoEngineered MAterials and Surfaces and
Dipartimento di Chimica, Materiali e Ingegneria Chimica G. Natta,
Politecnico di Milano, P.zza Leonardo da Vinci 32 I-20133 Milano, Italy

**Abstract**

Linear sp carbon nanostructures are gathering interest for the physical properties of one-dimensional (1D) systems. At present, the main obstacle to the synthesis and study of these systems is their instability. Here we present a simple method to obtain a solid system where linear sp chains (i.e. polyynes) in a silver nanoparticle assembly display a long term stability at ambient conditions. The presence and the behavior of linear carbon is investigated by Surface Enhanced Raman Scattering (SERS) exploiting the plasmon resonance of the silver nanoparticles assembly. This model system opens the possibility to investigate an intriguing form of carbon nanostructures.

In recent years a renewed interest has been devoted to carbon nanostructures with sp hybridization which represent a real carbon nanowire and a one-dimensional (1D) conjugated carbon system [1,2]. Sp-hybridized carbon atoms are predicted to form linear chains with two possible electronic configurations: structures displaying single-triple alternating bonds are generally called polyynes while the name cumulenes usually refers to structures with cumulated double bonds [3].

Sp carbon structures have been produced or observed in many solid carbon-based systems such as carbon nanotubes [4], nanostructured cluster-assembled carbon films [5] and amorphous carbon [6]. Isolated sp chains have been also produced in solution by different chemical and physical methods [2, 7, 8] or alternatively deposited in vacuum and embedded in a cryogenic solid inert gas matrix [9, 10].

At the moment the stability of the sp phase in all the above mentioned systems is the main problem to deal with [11]. In fact chain terminations, chain-chain crosslinking reactions and exposure to oxygen play a fundamental role in determining structural reorganizations towards the more stable $sp^2$ phase [12]. Carbon chains (namely polyynes) terminated with hydrogen or nitrogen atoms are relatively stable as long as they are dissolved in suitable solvents acting as a separator; stability is easily lost at relatively high concentrations [13], under thermal treatments [14] or after exposure to light [2, 15].

We recently studied linear carbon structures (polyynes) in solution by Raman and Surface Enhance Raman Scattering (SERS) showing also the high SERS sensitivity to a very low amount of sp structures when silver colloids are used as the SERS active medium [16]. Here we propose the use of silver nanoparticles to stabilize the chains by simply mixing with silver colloids. Evaporation of the solvent leads to a solid state nanostructured deposit where sp chains can survive in ambient atmosphere for long times (months), as attested by SERS measurements.

Hydrogen terminated linear carbon chains with polyyne structure were produced by electric arc discharge between two graphite electrodes submerged in methanol (further details can be found in [17]).

Silver colloids dispersed in water (average size about 40-60 nm [18]), synthesized following the method of Lee and Meisel [19], were added to the polyyne solution and then deposited on glass or silicon substrates. The solid deposits have been characterized by SERS and Scanning Electron Microscopy (SEM) with a Jobyn-Yvon Labram HR800 using the 785 nm wavelength of a solid state laser and with a Zeiss Supra 40 field emission scanning electron microscope (FE-SEM), respectively.

We already reported a Raman and SERS investigation of H-terminated polyynes in methanol solution [16]. Features are clearly visible in the 1800-2200 cm$^{-1}$ spectral range, where Raman active vibrational modes of sp structures are predicted [20]. On the basis of DFT calculations, modes in the 2000-2200 cm$^{-1}$ region were associated to shorter polyynes (<12 carbon atoms) while the 1800-2000 cm$^{-1}$ region can be related to longer chains (>12 carbon atoms), even if the presence of cumulenic chains cannot be *a priori* excluded [16].

The SERS spectra of the dried solid deposit (i.e. after solvent evaporation) and of polyynes in solution are reported in fig. 1. The spectrum of a Ag nanoparticle solid assembly without carbon chains is also reported for reference. The features in the 1800 -2200 cm$^{-1}$ range are substantially the same in solution and in the solid state assembly even though there are small differences in the relative intensity of some peaks (see inset fig.1). Features in the 1300-1600 cm$^{-1}$ spectral range are related to the presence of sp$^2$ coordinated amorphous carbon [21]. The increasing intensity of these features in the dried sample (as prepared and after 3 months) cannot be completely ascribed to reorganization of linear structures. In fact SERS on the dried samples is also sensitive to carbon contaminations coming from the atmosphere as revealed in the spectrum of a dried Ag nanoparticle sample without carbon chains [22]. SEM images of the dried material deposited on silicon substrates, as reported in fig.2, reveal a nanostructured assembly of silver nanoparticles of different shapes (comprising also elongated structures), with a diameter of about 60-80 nm. This occurrence permits to exploit the SERS effect also in the solid system to study polyyne behavior.

In order to test the stability of linear structures in the Ag assembly, we left this sample in air, just protecting it from the environmental dust and light. After about 3 months the SERS spectrum (reported in fig. 1) clearly shows that polyyne chains are still present and this evidence suggests an enhanced stability in the solid state sample. In any case if we compare more carefully the SERS spectrum after 3 months with the SERS spectrum of the as prepared solid sample, we notice a reduction of the intensity in the 1800-2000 $cm^{-1}$ region and an increase in the 2000-2200 $cm^{-1}$ region. This behavior points to a different stability of the short carbon chains with respect to the longer ones. In particular, the most stable structures are those associated to the bands around 2100 $cm^{-1}$ which were present in the pristine solution and correspond to polyyne chains of 8 and 10 carbon atoms [16]. Such structures represent the maximum in the chain length distribution in polyyne solutions produced by different techniques, therefore they can be expected to be the most stable also in the Ag solid assembly [8,17].

Our observations confirm that 1D carbon structures are stabilized by termination with silver nanoparticles and that a solid Ag nanoparticle assembly containing stable sp carbon structures can be obtained. Metals are already known as stabilizing agents for the sp phase: for instance Kavan and co-workers proposed chemical methods to produce $sp^2$-sp moieties where alkali atoms catalyze the formation of sp carbon and then act as separators and prevent cross linking reactions [23]. Also, Ag films evaporated on PTFE were already demonstrated to be able to produce sp structures under visible laser irradiation [25]. Therefore it is not surprising that silver may form complexes with linear carbon chains (a similar behavior has been observed with copper [24]). More surprising is instead the stability enhancement observed in our system and this effect could be ascribed to the same interaction mechanism between an sp carbon chain and a silver nanoparticle which enables the SERS chemical effect. In other words a bonding with silver may help the sp chain to stabilize its electronic configuration, thus preventing a reorganization pointing to a more stable $sp^2$ configuration.

The main advantage of our system is the fully knowledge of the linear carbon structures inserted in the colloidal solution, while in other sp containing solid systems such as sp-rich amorphous carbon [4-6] the chain length distribution and the chain termination are substantially unknown. Isolated sp structures in solution can be produced controlling the terminating atom (e.g. hydrogen or C-N group) and with a carefully characterized chain length distribution (measured by high performance liquid chromatography HPLC). This control allows to study physical and chemical properties of linear carbon as a function of chain length (i.e. number of carbon atoms), since the confinement effects and the structure-related properties in such 1D conjugated pure carbon nanowires are presently at the center of interest in physics, chemistry and material science [2].

In conclusion we here demonstrated the capability to obtain a solid system containing stable linear carbon structures by simply drying a solution of sp carbon chains mixed with silver colloids. Sp chains can survive in this system for several months without undergoing substantial modifications. SERS measurements permit to detect isolated sp chains and to follow their evolution after the drying process. This material represents a model system (with a well known chain length distribution) to investigate the properties and the behavior of isolated sp structures also by other characterization techniques not suitable for solutions. Moreover, from a technological point of view it could be a good candidate for optical and electronic applications owing to the electronic transport properties [26] and to the nonlinear optical properties theoretically predicted and recently experimentally detected [27].

**Acknowledgments**

The authors acknowledge A. Bonetti and S. Salvatore for the contribution given during their undergraduated thesis project.

**References**

[1]     S. Szafert, J. A. Gladysz, *Chem. Rev.* **103**, 4175-4205 (2003).

[2]     F. Cataldo (editor), *"Polyynes: Synthesis, Properties and Applications"*, CRC press, Taylor & Francis publishing group, Boca Raton, (2005)

[3]     R. B. Heimann, S. E. Evsyukov and L. Kavan (editors) *"Carbyne and carbynoid structures"*, Kluwer Academic Publishers (1999).

[4]     X. Zhao, Y. Ando, Y. Liu, M. Jinno, T. Suzuki, *Phys. Rev. Lett.* **90**, 187401 (2003).

[5]     L. Ravagnan, F. Siviero, C. Lenardi, P. Piseri, E. Barborini, P. Milani, C.S. Casari, A. Li Bassi, C.E. Bottani, *Phys. Rev. Lett.*, **89**, 285506-1 (2002).

[6]     L. Kavan et al., Chem. Rev.  97, 3061-3082 (1997).

[7]     M. Tsuji et al., Carbon **41** (2003) 2141.

[8]     H. Tabata, M. Fuji and S. Hayashi, Chem. Phys. Lett. **395** (2004) 138.

[9]     T. Wakabayashi et al., J. Phys. Chem. **108** (2004) 3686.

[10]    D. Strelnikov, R. Reusch, and W. Kratschmer, J. Phys. Chem. A **2005,** 109, 7708-7713

[11]    R.H. Baughmann, Science 312 (2006) 1009

[12]    C.S. Casari, A. Li Bassi, L. Ravagnan, F. Siviero, C. Lenardi, P. Piseri, G. Bongiorno, C.E. Bottani, P. Milani, *Phys. Rev. B*, **69** (2004) 075422.

[13]    F. Cataldo, Polym. Degrad. Stabil. **91** (2006) 317

[14]    D. Heimann, Carbon **43** (2005) 2235–2242

[15]    F. Cataldo, Fullerenes, Nanot. Carbon Nanostruct. 12 (2004) 633.

[16]     A. Lucotti, M. Tommasini, M. Del Zoppo, C. Castiglioni, G. Zerbi, F. Cataldo, C.S. Casari, A. Li Bassi, V. Russo, M. Bogana, C.E. Bottani *Chem. Phys. Lett.* **417**, 78-82 (2006).

[17]  F.Cataldo, Carbon, **42** (2004) 129; F. Cataldo, Tetrahedron **60** (2004) 4265.

[18]  P. Kamat, M. Flumiani, G. Hartland, J. Phys. Chem. B, **102** (1998) 3123.

[19]  P. C. Lee and D. J. Meisel, J. Phys. Chem., **86** (1982) 3391.

[20]  J. Kürti et al., Synth. Met. **71** (1995) 1865.

[21]  A.C. Ferrari and J. Robertson, *Phys. Rev. B* **61** (2000) 14095.

[22]  A. Otto, *J. Raman Spectrosc.* 2002; **33**: 593–598

[23]  J. Kastner et al., Macromol. **28** (1995) 344.

[24]  F.Cataldo, Carbon 43 (2005) 2792

[25]  L. Ravagnan, F. Siviero, C.S. Casari, A. Li Bassi et al., Carbon 43 (2005) 1317–1339

[26]  S. Tongay, R.T. Senger, S. Dag, and S. Ciraci, Phys. Rev. Lett. 93, 136404-1 (2004)

[27] D. Slepkov, F. A. Hegmann, S. Eisler, E. Elliott, R. R. Tykwinski, *J. Chem. Phys.* **120**, 6807-6810 (2004).

**Figure Captions**

**Figure 1**

SERS spectra of linear carbon in solution and in the solid Ag nanoparticle assembly as prepared and after 3 months. Spectra are normalized to the sp signal. The SERS spectrum of the Ag solid assembly without linear carbon is also reported for reference. Particular of the sp signal in the SERS spectra is reported in the inset.

**Figure 2**

SEM image of the Ag nanoparticle solid assembly deposited on a silicon substrate.

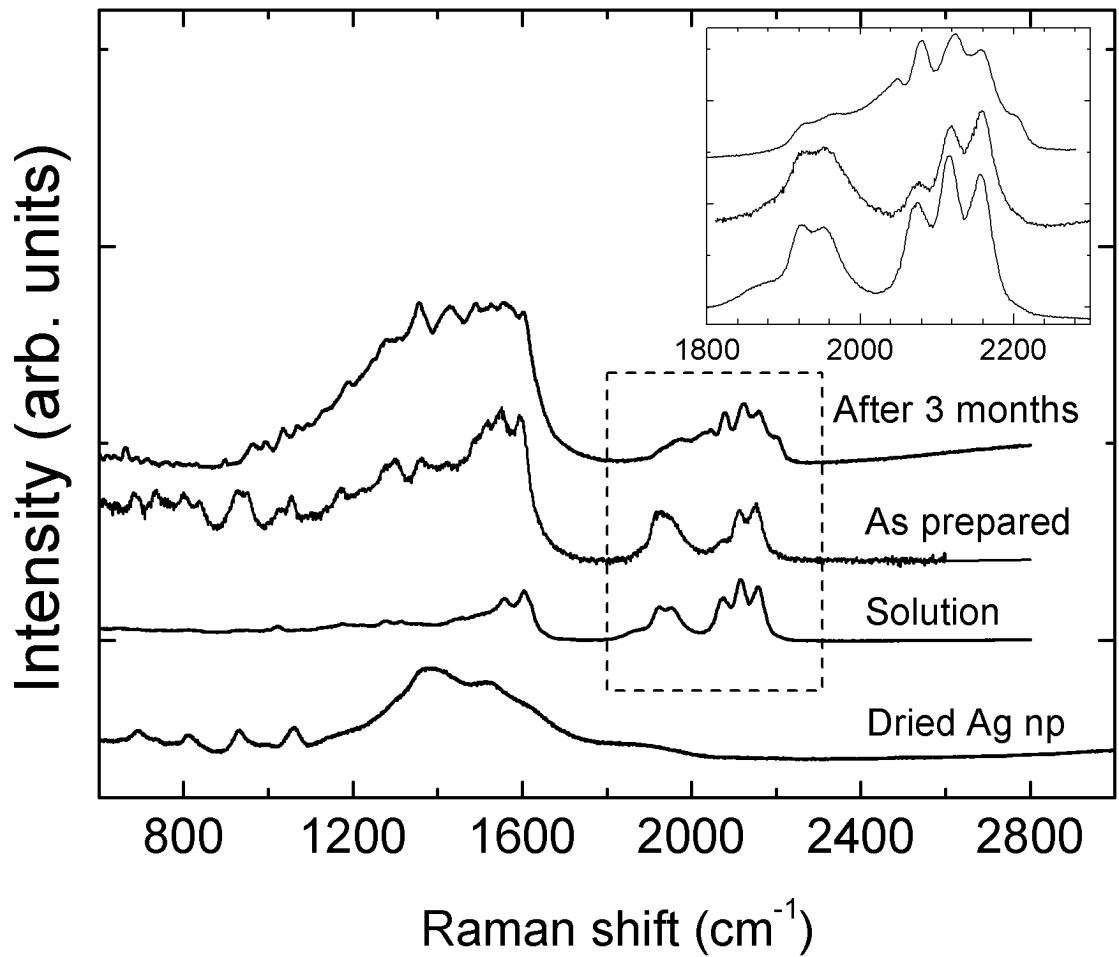

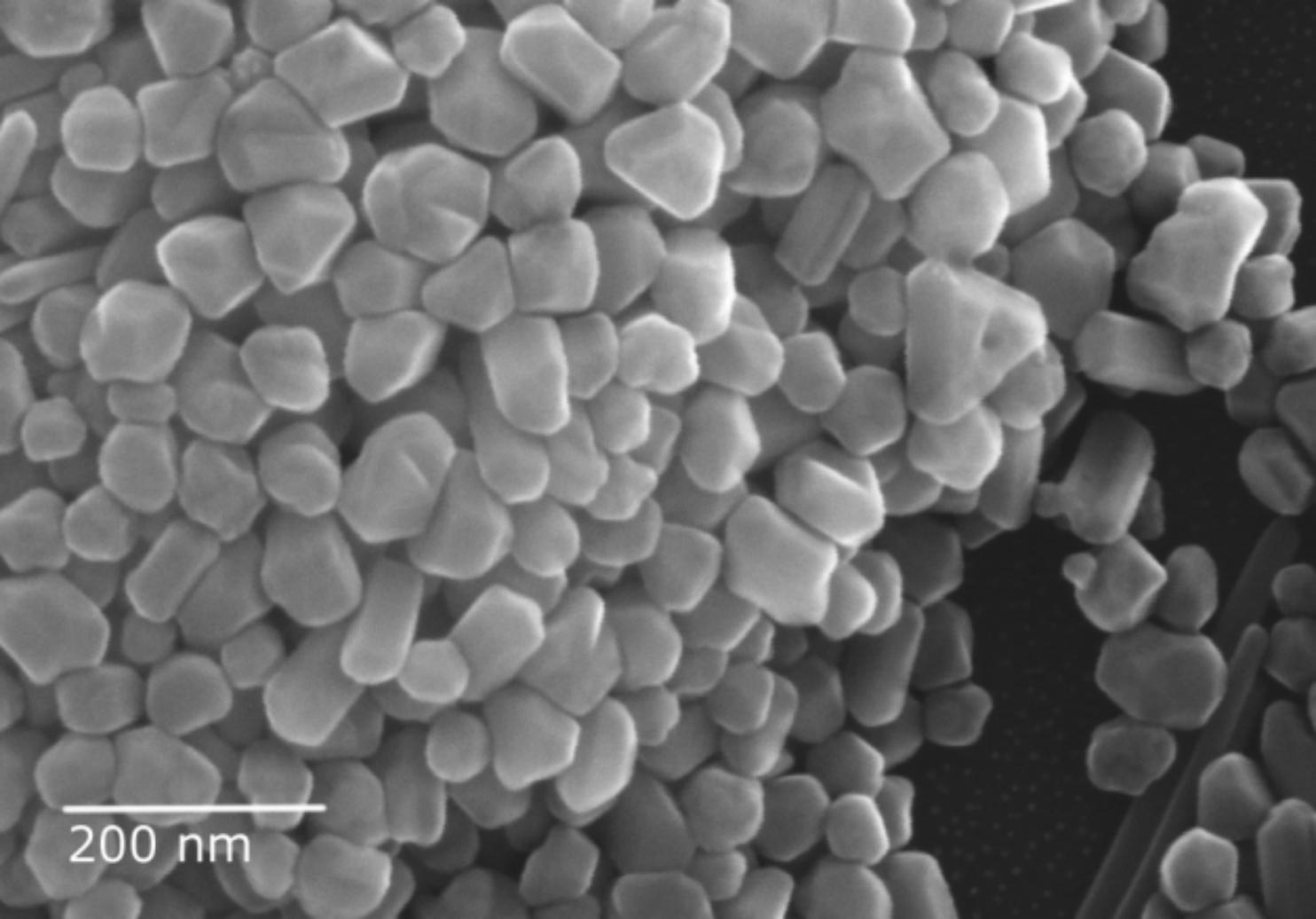